\documentclass[prl,twocolumn,superscriptaddress]{revtex4-1}

\usepackage[dvipdfmx]{graphicx}
\usepackage{dcolumn}% Align table columns on decimal point
\usepackage{bm}% bold math
\usepackage{color}
\newcommand{\Vec}[1]{\mbox{\boldmath$#1$}}
\graphicspath{{fig/}}

\begin{document}

\preprint{APS/123-QED}

\title{
Possible counterintuitive enhancement of superconductivity in ladder-type cuprates by longitudinal compression 
} 
\author{Hikaru Sakamoto}
%\affiliation{Department of Physics, Osaka University, 1-1 Machikaneyama-cho, Toyonaka, Osaka, 560-0043, Japan}
\author{Kazuhiko Kuroki}
\email{Corresponding author}
\affiliation{Department of Physics, Osaka University, 1-1 Machikaneyama-cho, Toyonaka, Osaka, 560-0043, Japan}

\date{\today}% It is always \today, today,

\begin{abstract}
  We theoretically study the effect of uniaxial deformation of ladder-type cuprate superconductors. Model construction based on first principles calculation shows that the rung-to-leg ratio of the nearest neighbor hoppings counterintuitively increases when the lattice is compressed in the longitudinal (leg) direction. This leads to an enhancement of the superconducting transition temperature, which intuitively is expected when compressed in the rung direction. Such a trend is traced back to the on-site hybridization between Cu$4s$ and Cu$d_{x^2-y^2}$ orbitals, which varies and changes sign upon lattice deformation.
\end{abstract}

\pacs{ }% PACS, the Physics and Astronomy aiueofdsfs
\maketitle

Superconductivity in ladder-type cuprates has been studied extensively both theoretically and experimentally following the seminal proposal by Dagotto and Rice. \cite{Dagotto,Rice,DagottoRice} In fact, (Sr,Ca)$_{14}$Cu$_{24}$O$_{41}$ compound\cite{Review142441}, which consists of two-leg ladders and chains, were found to be superconducting with a $T_c$ of above 10K under high pressure\cite{Uehara}. Theoretically, it was suggested that a stronger spin-spin coupling in the rung direction enhances the superconducting transition temperature\cite{Dagotto2}. Intuitively, it is expected that a shorter Cu-Cu distance in the rung direction would result in a larger nearest neighbor electron hopping amplitude in that direction, and hence stronger spin-spin coupling within the rung. 
Nowadays, there is a renewed interest in the problem of ladder type materials since a two-leg ladder lattice can be viewed as a two-band system where wide and narrow bands coexist\cite{Kuroki,Ogura,Matsumoto,OguraDthesis}. In such a system, when the Fermi level is placed in the vicinity of the narrow band edge, strong enhancement of superconductivity is expected. 

In the present study, we explore how the electron hoppings of the ladder type cuprates are affected when uniaxial compression or tension is applied to the lattice in the leg or rung directions, and investigate its consequences to superconductivity. For simplicity, we consider the two-leg ladder cuprate without the chains, that is, SrCu$_2$O$_3$\cite{Hiroi}, although this material is known to be difficult to dope carriers. We surprisingly find that the ratio $t_r/t_l$, where $t_r (t_l)$ is the nearest neighbor hopping in the rung (leg) direction, is enhanced when the lattice is compressed in the leg direction or stretched in the rung direction. This counterintuitive manner of the hopping variation can be attributed to the on-site hybridization between Cu $d_{x^2-y^2}$ and Cu $4s$ orbitals, which arises due to the low symmetry of the lattice\cite{Anisimov}. Due to such variation of the hoppings, we find that superconducting transition temperature ($T_c$) is enhanced when the lattice is compressed in the leg direction, opposed to an intuitive expectation. The effect is expected to be strong especially in the electron-doped regime. 

The model construction of SrCu$_2$O$_3$ is performed as follows.
We take the lattice constant determined experimentally\cite{structexp} as a reference, and assume crystal structures compressed or stretched by certain amount in the leg or rung directions. We determine the internal coordinates for these crystal structures through structural optimization and calculate the electronic band structure, using the Perdew-Burke-Ernzerhof parametrization of the generalized gradient approximation (PBE-GGA)\cite{PBE-GGA} and the projector augmented wave method\cite{Kresse} as implemented
in the VASP code\cite{VASP1,VASP2,VASP3,VASP4}. Plane-wave cutoff energy and the $k$-meshes were taken as 550eV and $10\times 10\times 10$, respectively.
We then extract the Wannier functions\cite{Marzari,Souza} from the calculated band structures using the WANNIER90\cite{Wannier90} code, which gives the tightbinding hoppings and on-site energies $t_i^{\alpha\beta}$, where $i$ and $\alpha,\beta$ denote the lattice vectors and the orbitals, respectively. The tightbinding model in momentum space is obtained in the form $\varepsilon_{\alpha\beta}(\Vec{k})=\sum_i^N t_i^{\alpha\beta} \exp(i\Vec{k}\cdot\Delta\Vec{r}_i)$, where we take $N=621$ lattice vectors $\Delta\Vec{r}_i$.
To the obtained tightbinding model, we add the on-site interaction $U$ term, and the many-body study is performed within the fluctuation exchange approximation (FLEX)\cite{Bickers}. 
We obtain the renormalized Green's function by solving the Dyson's equation in a self-consistent calculation.  
The obtained Green's function and the pairing interaction mediated mainly by spin-fluctuations 
are plugged into the linearized Eliashberg equation. 
The superconducting transition temperature $T_c$ is determined as the temperature where the eigenvalue of the Eliashberg equation reaches unity\cite{comment}. In the FLEX calculation, $32\times 32\times 4$ $(k_x,k_y,k_z)$-meshes were taken.

\begin{figure}
	\includegraphics[width=9cm]{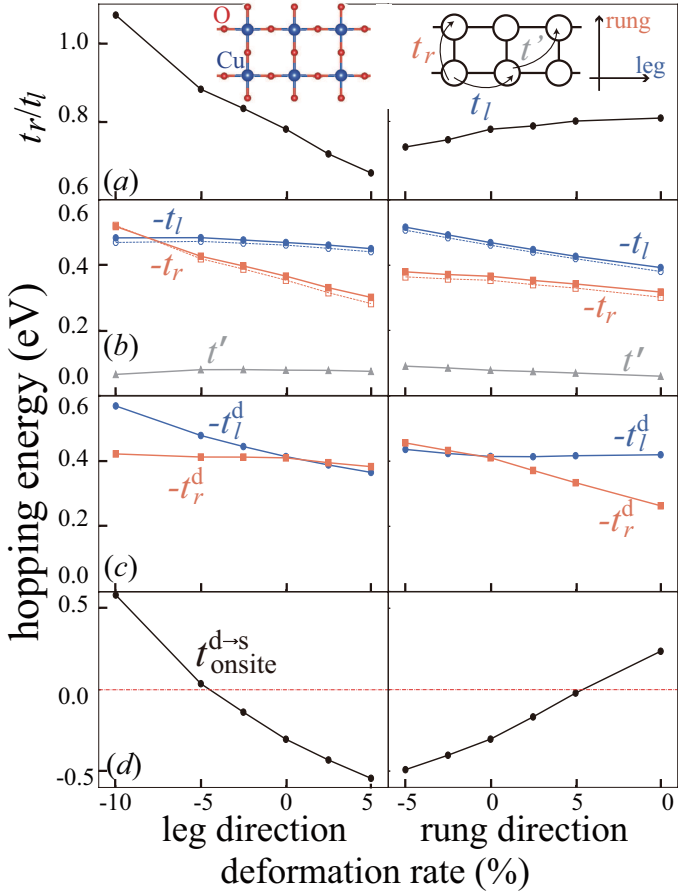}
	\caption{The variation of the hoppings against uniaxial deformation in the leg (left panels) or rung (right panels) directions. We take the tension (+) or compression (-) rate as the horizontal axis. (a) The ratio $t_r/t_l$. (b) $t_l$, $t_r$, and $t'$ of the two-orbital model (solid lines) and the four-orbital estimation (see text) of $t_l$ and $t_r$ (dashed lines). (c) $t_l^d$ and $t_r^d$ in the four-orbital model. (d) $t_{\rm onsite}^{d\rightarrow s}$ in the four-orbital model.}
	\label{fig1}
\end{figure}

Here, we first construct a model where we explicitly consider the $d_{x^2-y^2}$ orbital centered at the Cu site. This will be referred to as the two-orbital model since there are two Cu sites per unit cell (this ``$d_{x^2-y^2}$'' Wannier orbital consists of a mixture of Cu $d_{x^2-y^2}$, oxygen $2p$, and also, as explained later, Cu $4s$ atomic orbitals). From this model, we estimate the nearest neighbor hoppings in the leg ($t_l$) and rung ($t_r$) directions, and also the next nearest neighbor diagonal hopping $t'$ (see the inset of Fig.\ref{fig1}(a)).
In Fig.\ref{fig1}(b)(solid lines), we present the variation of the hoppings $t_l$, $t_r$, and $t'$ upon compressing or stretching the lattice in the leg or rung directions, and in Fig.\ref{fig1}(a) the variation of the ratio $t_r/t_l$. In contrast to an intuitive expectation, $t_r/t_l$ increases when the lattice is compressed in the leg direction and stretched in the rung direction.

\begin{figure}
	\includegraphics[width=9cm]{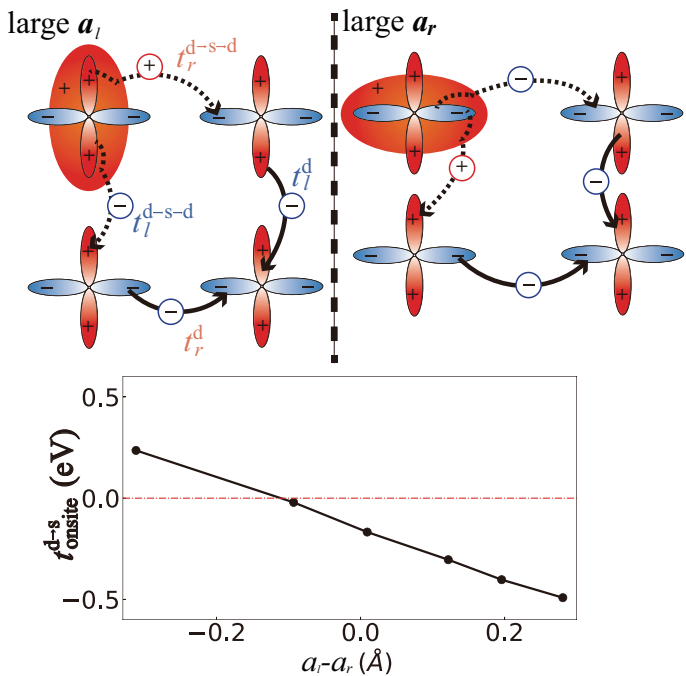}
	\caption{Upper panel : schematic image of the hoppings of the four orbital model in the large $a_l$ (left) and large $a_r$ (right cases). Lower panel : The on-site hopping from Cu$4s$ to Cu$d_{x^2-y^2}$ orbitals in the four orbital model, plotted against the difference between the nearest neighbor distances in the leg and rung directions.}
	\label{fig2}
\end{figure}

To understand the origin of this counterintuitive variation of the hoppings against the lattice deformation, we now construct a model which explicitly takes into account the Cu $3d_{x^2-y^2}$ and $4s$ orbitals. This model will be called the four-orbital model. In fact, it has been known that the $4s$ orbital hybridizes with $d_{x^2-y^2}$ to give rise to an appreciable diagonal hopping in the cuprates\cite{Andersen,Pavarini,SakakibaraPRB,SakakibaraPRL}. For the ladder structure in particular, it was pointed out in ref.\cite{Anisimov} that the anisotropy of the $4s$-orbital-related hoppings is the origin of the anisotropy of the effective $d$-$d$ hoppings in the leg and rung directions.  Note that the effect of the $4s$ orbital is implicitly taken into account in the Wannier orbitals in the two-orbital model. In Fig.\ref{fig1}(c), we plot the hopping between the nearest neighbor $d_{x^2-y^2}$ orbitals in the leg ($t_l^d$) and rung ($t_r^d$) directions\cite{comment2}. Now these hoppings behave as intuitively expected, namely, $t_l^d$ becomes large when the lattice is compressed in the leg direction, and $t_r^d$ is reduced when the lattice is stretched in the rung direction. Similarly, we find that the nearest neighbor hoppings between $4s$ and $d_{x^2-y^2}$ ($t_{l,r}^{s\rightarrow d}$) behave as intuitively expected under lattice deformation (not shown). We therefore expect that the hoppings between the $d_{x^2-y^2}$ orbitals via the $4s$ orbital (see Fig.\ref{fig2}) play a crucial role in the counterintuitive lattice deformation dependence of the hoppings in the two-orbital model. We estimate the $d\rightarrow s\rightarrow d$ hopping using second order perturbation theory as 
\begin{equation}
t_{l,r}^{d\rightarrow s \rightarrow d}=\frac{t_{\rm onsite}^{d\rightarrow s}t_{l,r}^{s\rightarrow d}}{\varepsilon_d-\varepsilon_s},
\label{eq1}
\end{equation}
where 
$t_{\rm onsite}^{d\rightarrow s}$ is the $d_{x^2-y^2}$ to $4s$ hopping within the same Cu site,  $t_{l,r}^{s\rightarrow d}$ is the nearest neighbor $4s$ to $d_{x^2-y^2}$ hopping in the leg or rung directions, and $\varepsilon_{d,s}$ is the on-site energy of the $d_{x^2-y^2}$ or $4s$ orbitals. Contribution from all possible equivalent paths are added up, and added to $t_{l,r}^{d}$, which gives the dashed line plots in Fig.\ref{fig2}. As seen in this plot, the four-orbital estimation almost perfectly reproduces the two-orbital results, which confirms the view that the origin of the counterintuitive variation of $t_r$ and $t_l$ is the hopping path $d\rightarrow s\rightarrow d$ (we have also checked that contributions coming from other paths that involve the $4s$ orbital are very small).

To further understand intuitively the contribution from the $d\rightarrow s\rightarrow d$ path, we focus on $t_{\rm onsite}^{d\rightarrow s}$, plotted in Fig.\ref{fig1}(d), which changes sign upon lattice deformation. This sign change can be intuitively understood from the upper panels of Fig.\ref{fig2}. Namely, when the nearest neighbor Cu-Cu distance in the leg direction $a_l$ is long, the widely spread $4s$ orbital is elongated in the leg direction, while the more localized $d_{x^2-y^2}$ orbital is less deformed. In this case, the on-site hopping is dominated by the longitudinal portion of the $d_{x^2-y^2}$ wave function, so that $t_{\rm onsite}^{d\rightarrow s}<0$ , taking the phase of the orbitals as depicted in the figure (note that the sign of a hopping is the opposite to that of the  multiplication of the signs of the wavefunction of the initial and final orbitals). Similarly, when the nearest neighbor Cu-Cu distance in the rung direction $a_r$ is long and the $4s$ orbital is elongated in the rung direction, $t_{\rm onsite}^{d\rightarrow s}>0$. This tendency is confirmed in the calculation result shown in the lower panel of Fig.\ref{fig2}. When $a_l$ is large and hence $t_{\rm onsite}^{d\rightarrow s}<0$, from eqn.(\ref{eq1}), $t_{r}^{d\rightarrow s \rightarrow d}>0$ and $t_{l}^{d\rightarrow s \rightarrow d}<0$ because $t_{r}^{s\rightarrow d}>0$, $t_{l}^{s\rightarrow d}<0$, and $\varepsilon_d-\varepsilon_s<0$. Similarly, when $a_r$ is large and hence $t_{\rm onsite}^{d\rightarrow s}>0$, the sign of the indirect hoppings becomes the opposite as $t_{r}^{d\rightarrow s \rightarrow d}<0$ and $t_{l}^{d\rightarrow s \rightarrow d}>0$. Adding $t_{l,r}^{d\rightarrow s \rightarrow d}$ to the negative $t_{l,r}^d$ explains the counterintuitive variation of the hoppings against the lattice deformation.

\begin{figure}
	\includegraphics[width=9cm]{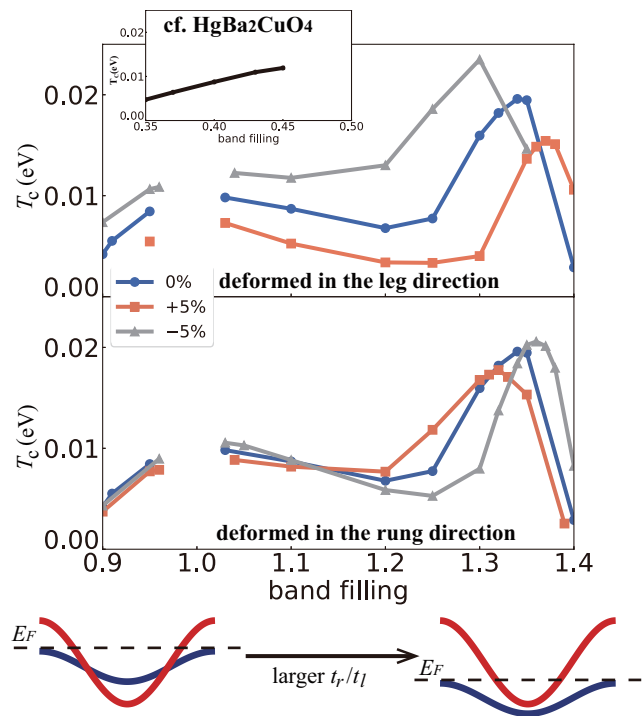}
	\caption{Upper panels : The FLEX result of the superconducting transition temperature calculated for the two-orbital model against lattice deformation in the leg (upper) or rung (lower) directions. Inset: A similar calculation result for the single orbital model of HgBa$_2$CuO$_4$. Lower panel : A schematic image of the relative shift of the bonding and antibonding bands as $t_r/t_l$ increases.}
	\label{fig3}
\end{figure}

We now move on to the analysis of superconductivity. We take the on-site repulsion $U=3$eV, which is a typical value for the cuprates\cite{Han}. In the upper panels of Fig.\ref{fig3}, we plot the superconducting transition temperature against the band filling $n$(=number of electrons/number of sites) for the cases when the lattice is compressed or stretched by $5\%$ in the leg or rung directions. A common feature in all cases is the double local maximum of $T_c$, one around half filling, and another at $n>1$, i.e., in the electron doped regime. $T_c$ is enhanced near half-filling due to the enhancement of electron correlation. A prominent feature peculiar to the ladder-type lattice is the rather high $T_c$ in the electron-doped regime. The $T_c$ maximum in this regime is about twice as high as that of the $T_c$ calculated in the same way for a $100$K cuprate superconductor HgBa$_2$CuO$_4$, shown in the inset of Fig.\ref{fig3}. To understand this $T_c$ maximum, we introduce the tightbinding band dispersion of the two-leg ladder given as 
$ %\begin{equation}
E_\pm(k)=t_l\left[2\left(1\mp \frac{t'}{t_l}\right)\cos(k)\pm \frac{t_r}{t_l}\right]
$, %\end{equation}
where $-$ and $+$ stand for bonding and antibonding bands, respectively, and $-1<t'/t_l<0$ (appropriate for the cuprates) makes the bonding band narrower than the antibonding band. In the previous studies\cite{Kuroki,Matsumoto}, it was shown that $T_c$ is strongly enhanced when the Fermi level is raised by electron doping (the necessity of about 30 percent electron doping was suggested in Ref.\onlinecite{Kuroki}) so that it lies just above the top of the bonding band. There, this was considered as an example of superconductivity enhanced in systems with coexisting wide and narrow bands, when the Fermi level is positioned in the vicinity of the narrow band edge, namely, when the narrow band is ``incipient''\cite{DHLee,Hirschfeld,Hirschfeldrev,YBang,YBang2,YBang3,Borisenko,Ding,MaierScalapino2}. 

Now, when the lattice is compressed in the leg direction or stretched in the rung direction, the ratio $t_r/t_l$ counterintuitively increases (while $t'/t_l$ is barely affected) as we have seen, so that the bonding band is lowered relatively to the antibonding band, as depicted schematically in the lower panel of Fig.\ref{fig3}. Hence, less electron is required for the Fermi level to reach the vicinity of the bonding band top. This is the reason why the $T_c$ maximum moves toward the less-electron-doped regime in these cases. Especially when the lattice is compressed in the leg direction, the maximum $T_c$ itself is enhanced  because the electron correlation effect becomes stronger as the band filling approaches half filling. On the other hand, when the lattice is stretched in the rung direction, although $T_c$ is maximized in the less-electron-doped regime, the maximum $T_c$ is suppressed. We believe this is because $|t_l|$ decreases, leading to the reduction of the energy scale.

\begin{figure}
	\includegraphics[width=9cm]{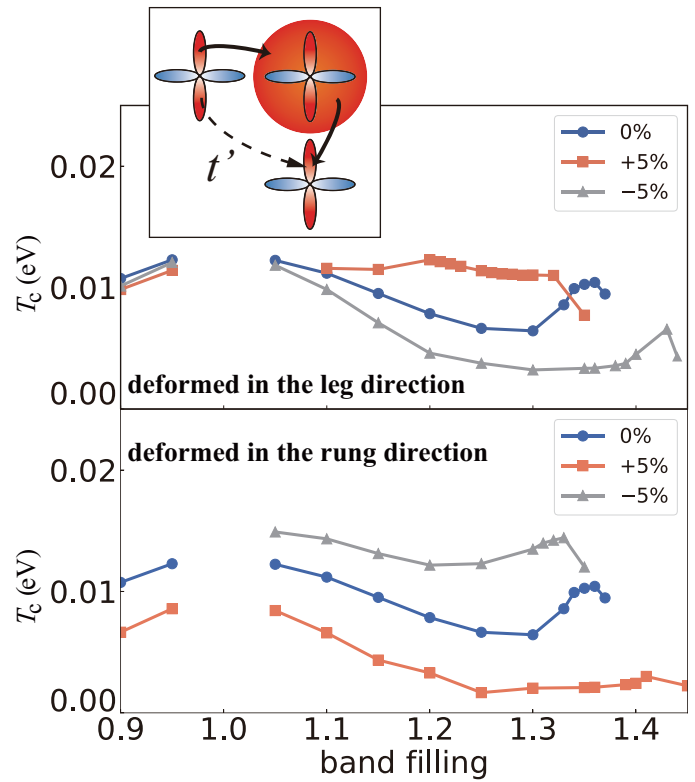}
	\caption{Plots similar to Fig.\ref{fig3} for the ``$d$-only'' two orbital model, obtained by removing the $4s$ orbitals from the four orbital model. Inset : A schematic image of the origin of the diagonal hopping $t'$, which involves the $4s$ orbital.}
	\label{fig4}
\end{figure}

Let us now extract the $4s$ orbital effect that is implicitly taken into account in the two-orbital model. To do this, we consider a two-orbital model obtained by removing the $4s$ orbitals ``by hand'' from the four-orbital model. $T_c$ calculated for this ``$d$-only'' two-orbital model against the band filling for the same lattice deformation as in the original two-orbital model is shown in Fig.\ref{fig4}. The trend is almost completely the opposite compared to the original two-orbital model, namely, the local $T_c$ maximum is reduced and moves toward the more-electron-doped regime when the lattice is compressed in the leg direction or stretched in the rung direction. This is due to a combination of two effects that involve the $4s$ orbital. One is that the diagonal hopping $t'$ becomes very small in the absence of $4s$ because the main origin of $t'$ is the hopping path via the $4s$ orbital, as shown schematically in the inset of Fig.\ref{fig4}\cite{Andersen,Pavarini,SakakibaraPRB,SakakibaraPRL}. In such a case, the bonding and antibonding bands have nearly the same band width (and the band structure is nearly electron-hole symmetric). This requires more amount of electron doping for the Fermi level to reach the top of the bonding band. Hence, the local maximum $T_c$ is suppressed compared to the original two-orbital (see the supplemental material for details). Another effect is that the counterintuitive variation of $t_r/t_l$ is lost in the absence of $4s$. Therefore, compressing the lattice in the leg direction or stretching it in the rung direction simply suppresses $t_r/t_l$, requiring large amount of doped electrons for the bonding band to be incipient. These results conversely reveal the crucial role played by the implicitly considered $4s$ orbital in the original two-orbital model.

The analysis on superconductivity in our study is based on FLEX, which is kind of a weak-coupling approach, but the tendency of enhanced superconductivity with larger $t_r$ has also been pointed out in a previous density matrix renormalization group study\cite{Noack}. We expect that the main conclusion here is qualitatively unaffected even if we adopt a strong-coupling viewpoint, where the spin-spin couplings in the leg and rung directions are given as $J_l=4t_l^2/U$ and $J_r=4t_r^2/U$, respectively, in the large $U$ limit. The compression of the lattice in the leg direction enhances $t_r$ and hence $J_r$ (while keeping $t_l$ and $J_l$ almost unchanged), from which we expect superconductivity to be enhanced based on the previous studies on the $t$-$J$ model on ladder-type lattice\cite{Dagotto2,Troyer,Tsunetsugu,Sano,Hayward}. Although we are not sure about how the ``incipient band'' situation affects superconductivity in the $t$-$J$ model, we do expect larger effect in the electron-doped regime than in the hole-doped regime due to the larger density of states in the former.

Relevance of the present study to experiments is of great interest. In fact, it was found in ref.\cite{Uniaxial} that applying uniaxial pressure to (Sr,Ca)$_{14}$Cu$_{24}$O$_{41}$ in the leg direction results in an enhancement of $T_c$ compared to the case when hydrostatic pressure is applied\cite{Uehara,Nagata}, although the quantitative correspondence between theory and experiment is not clear, partially because this is the case when holes are doped in the ladder. 

To summarize, we have investigated how superconductivity in the ladder-type cuprates is affected through modification of the electronic structure when uniaxial compression or tension is applied. It is found that the ratio $t_r/t_l$ is enhanced when the lattice is compressed in the leg direction or stretched in the rung direction. This counterintuitive manner of the hopping variation is attributed to the on-site hybridization between Cu $d_{x^2-y^2}$ and Cu $4s$ orbitals, which varies as the $4s$ orbital is deformed through the lattice deformation. Due to such variation of the hoppings, $T_c$ is enhanced when the lattice is compressed in the leg direction, opposed to an intuitive expectation. The effect is expected to be strong especially in the electron-doped regime, where the Fermi level approaches the top of the bonding band.

\begin{acknowledgments}
We acknowledge Daisuke Ogura, Karin Matsumoto, Masayuki Ochi, and Hidetomo Usui for valuable discussions. This study is supported by JSPS KAKENHI Grant Number JP18H01860.
\end{acknowledgments}

%\bibliography{lanio2}

\begin{thebibliography}{99}
\bibitem{comment} FLEX calculation for the Hubbard model on ladder-type lattice is often performed for a two-dimensional lattice where the ladders are weakly coupled (see, e.g., ref.\cite{Matsumoto} and references therein). Since $T_c$ is barely affected by this weak ladder-ladder coupling, here we perform the FLEX calculation for a purely one-dimensional lattice.
\bibitem{comment2} Since the oxygen $2p$ orbitals are still implicitly considered in the four orbital model,  the $d_{x^2-y^2}$ Wannier orbital mainly consists of a mixture of Cu $d_{x^2-y^2}$ and O $2p$ atomic orbitals. Therefore, in terms of the atomic orbitals, the hoppings $t_l^d$ and $t_r^d$ here mainly originate from the path Cu $d_{x^2-y^2}$ $\rightarrow$ O $2p$ $\rightarrow$ Cu $d_{x^2-y^2}$.
\end{thebibliography}

\end{document}

\preprint{APS/123-QED}

\title{
Supplemental material: Possible counterintuitive enhancement of superconductivity in ladder-type cuprates by longitudinal compression
} 
\author{Hikaru Sakamoto}
%\affiliation{Department of Physics, Osaka University, 1-1 Machikaneyama-cho, Toyonaka, Osaka, 560-0043, Japan}
\author{Kazuhiko Kuroki}
\affiliation{Department of Physics, Osaka University, 1-1 Machikaneyama-cho, Toyonaka, Osaka, 560-0043, Japan}

\date{\today}% It is always \today, today,

\pacs{ }% PACS, the Physics and Astronomy aiueofdsfs
\maketitle

\setcounter{equation}{0}
\setcounter{figure}{0}
\setcounter{table}{0}
\setcounter{page}{1}

\renewcommand{\theequation}{S\arabic{equation}}
\renewcommand{\thefigure}{S\arabic{figure}}
\renewcommand{\bibnumfmt}[1]{[S#1]}
\renewcommand{\citenumfont}[1]{S#1}
\renewcommand{\thetable}{S\arabic{table}}

\begin{figure}[h]
	\includegraphics[width=8cm]{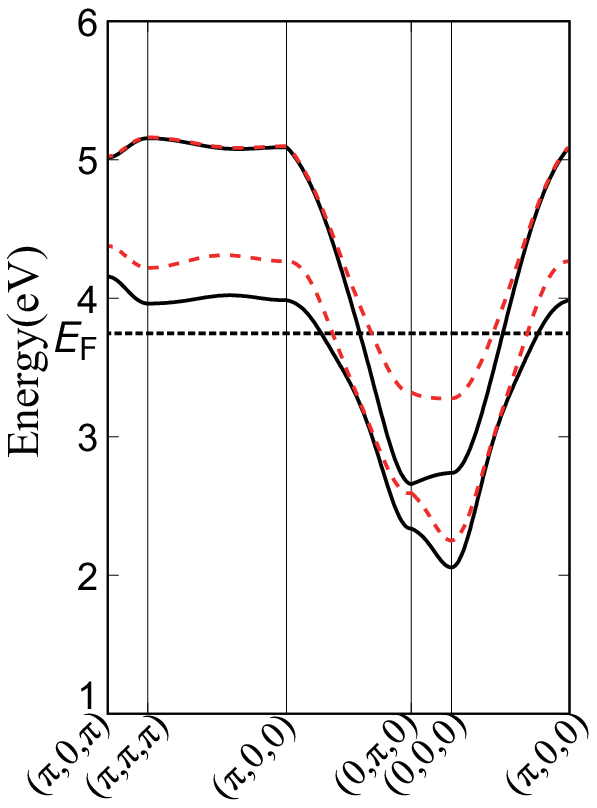}
	\caption{The band structure of the two-orbital model (black solid) and the $d$-only model (red dashed).}
	\label{figS1}
\end{figure}

\begin{figure}[h]
	\includegraphics[width=8cm]{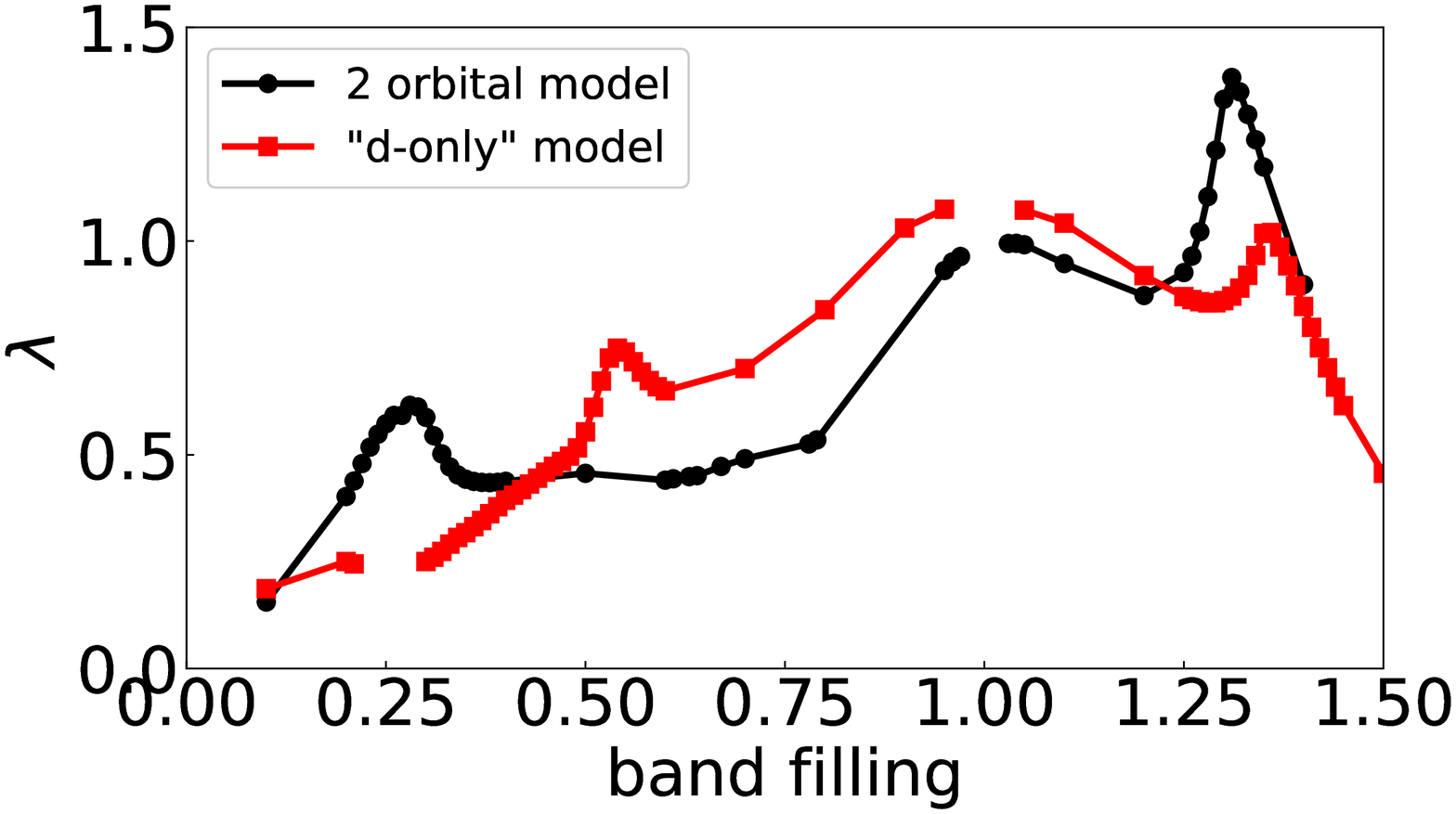}
	\caption{The comparison of the eigenvalue of the linearized Eliashberg equation at $T=0.01$eV as functions of the band filling.}
	\label{figS2}
\end{figure}

Here, we compare in more detail the two-orbital model and the $d$-only model. In Fig.\ref{figS1}, we compare the band structure. It can be seen that in the $d$-only model, the two bands have nearly the same band width, and the band structure is more electron-hole-symmetric. Therefore, larger band filling (more electrons) is required for the for the bonding band to become incipient. This means that superconductivity is locally optimized at a band filling which is further away from half filling compared to the case of the original two-orbital model, which results in a lower $T_c$ maximum. \ In Fig.\ref{figS2}, we plot the eigenvalue of the linearized Eliashberg equation at $T=0.01$eV as functions of the band filling for the two models. The result is more electron-hole-symmetric for the $d$-only model, as expected. The eigenvalue is maximized at a larger band filling for the $d$-only model, and hence its peak value is smaller than that for the original two-orbital model.

